\documentstyle[preprint,aps]{revtex}
\tightenlines
\begin{document}
\draft

\title{MAGNITOELASTIC INTERACTION AND LONG-RANG MAGNETIC ORDERING IN 
TWO-DIMENSIONAL FERROMAGNETICS }

\author{Yu.N. Mitsay, Yu.A. Fridman, D.V. Spirin, C.N. Alexeev, 
M.S. Kochma\'nski*\\}
\address{M.V.Frunze Simferopol State University,
Str. Yaltinskaya, 4, Simferopol, 333036, Ukraine.\\
*Institute of Physics, University of Rzesz\'ow, 
T.Rejtana 16 A, 35--310 Rzesz\'ow, Poland.\\
e-mail: mkochma@atena.univ.rzeszow.pl}
\date{\today}
\maketitle

\begin{abstract}
The influence of magnitoelastic (ME) interaction on the stabilization of 
long-range magnetic order (LMO) in the two-dimensional easy-plane 
ferromagnetic is investigated in this work. The account of ME exchange 
results in the root dispersion law of magnons and appearance of ME gap in the 
spectra of elementary excitations. Such a behavior of the spectra testifies 
to the stabilization of LMO and finite Curie's temperature. 
\end{abstract}
\pacs{PACS number(s): 75.10.-b}
\widetext

\section{Introduction}
It is well known $\cite{bloch30,pat82}$, that the existence of the long-range 
magnetic order (LMO) in $2D$ ferromagnetic is impossible. Formally it means, 
that in a Heisenberg ferromagnetic the integral 
\begin{eqnarray*}
\langle\Delta M\rangle\propto \int_0^{\omega}N(\omega)kdk,
\end{eqnarray*}
where $N(\omega)=[\exp(\omega/T)-1]^{-1}$, $\omega\propto k^2$ is the law 
of dispersion of spin waves, which determines fluctuations of the magnetic 
moment, diverges on the lower limit.

Berezinskii, Kosterlitz and Thouless $\cite{ber71,kost73}$ have shown, that in 
$2D$ easy-plane magnetics at finite temperatures there can exist only a 
quasilong-range order, that structure is determined rather by magnetization 
vortices, then by spin waves.

However in $\cite{mal76}$ it was shown, that the account of magnetodipole 
interaction in $2D$ ferromagnetics results in the root dispersion law of 
magnons $\omega\propto\sqrt{k}$. It causes the convergence of the integral 
$\langle\Delta M\rangle$ and testifies to stabilization of LMO at $T<T_c$, 
where $T_c$ is the temperature of the phase transition. Proceeding from this, 
it is possible to assume, that other relativistic types of interactions can 
stabilize LMO. For example, in paper $\cite{ivat96}$, it is shown, that in an 
easy-plane $2D$ antiferromagnetic LMO is stabilized by magnitoelastic (ME) 
interaction. However in this work there is used the Holstein-Primakoff 
representation for spin operators, i.e. quasiclassical representation, which 
is obviously inadequate to the microscopicity of the system in study.

Therefore it is of interest to investigate a question of stabilization of 
LMO in $2D$ easy-plane ferromagnetic with the exact account of one-ion 
anisotropy (OA) and ME interaction. Such an account can be made with the use 
of the Hubbard's operators procedure $\cite{zay75,mit89}$. This technique 
allows us to obtain the dispersion equation of hybridized magnitoelastic 
elementary excitations (bound ME waves), valid for arbitrary values of a 
magnetic ion spin ($S\geq 1$), temperatures (up to temperature of phase 
transition), and anisotropy constant. 

\section{Model system}
As a model system, we shall consider an easy-plane $2D$ ferromagnetic 
($XOZ$-basic plane), with the Hamiltonian:
\begin{eqnarray}
H=-\frac{1}{2}\sum_{n,n'}I(n-n')\vec{S}_n\vec{S}_{n'}+\frac{\beta}{2}\sum_n
(S^y_n)^2+\lambda\sum_n\{(S^x_n)^2u_{xx}+(S^z_n)^2u_{zz}+ \nonumber\\
(S^x_nS^z_n+S^z_nS^x_n)u_{xz}\}+
\int dv\frac{E}{2(1-\sigma^2)}[u^2_{xx}+u^2_{zz}+2\sigma u_{xx}u_{zz}+
2(1-\sigma)u^2_{xz}],
\end{eqnarray}
where $S^i_n$ is the spin operator in site $n$, $\beta >0$ is the OA constant, 
$\lambda$ is the constant of ME interaction, $u_{ij}$ is a symmetric part of 
tensor of deformations, $E$ is the modulus of elasticity, $\sigma$ is the 
Poisson coefficient. To simplify calculations we shall assume, that a magnetic 
ion spin $S=1$.

In the Hamiltonian (1) the first two terms describe the magnetic subsystem, 
the third is the two-dimensional ME interaction, and the fourth is the elastic 
subsystem.

As we study the possibility of stabilization of LMO due to the presence of 
ME exchange let us suppose that ME interaction creates the nonzero magnetic 
moment, which for definiteness, we consider parallel to $OZ$ axis. 

After separation of the mean field in the exchange part of (1), we obtain 
the one-ion Hamiltonian:
\begin{equation}
H_0(n)=-I_zS^z_n+\frac{\beta}{2}(S^y_n)^2+\lambda[u_{xx}(S^x_n)^2+u_{zz}
(S^z_n)^2+u_{xz}(S^x_nS^z_n+S^z_nS^x_n)].
\end{equation}
Solving the one-ion problem with the Hamiltonian (2) we find the energy 
levels of a magnetic ion:
\begin{equation}
\begin{array}{c}
E_1=\frac{\beta}{4}+\frac{\lambda}{2}(u^{(0)}_{xx}+2u^{(0)}_{zz})-\chi, \;\;\;
E_0=\frac{\beta}{2}+\lambda u^{(0)}_{xx}, \;\;\;
E_{-1}=\frac{\beta}{4}+\frac{\lambda}{2}(u^{(0)}_{xx}+2u^{(0)}_{zz})+\chi,\\
\chi^2=I^2_z+(\frac{\lambda}{2}u^{(0)}_{xx}-\frac{\beta}{4})^2, \;\;\;\;\;\;\;
I_z=I(0)<S^z>,
\end{array}
\end{equation}
and eigenfunctions of $H_0(n)$:
\begin{equation}
\Psi (1)=\cos\delta|1\rangle+\sin\delta|-1\rangle, \;\;\; \Psi(0)=|0\rangle, 
\;\;\; 
\Psi (-1)=-\sin\delta|1\rangle+\cos\delta|-1\rangle,
\end{equation}
where
\begin{eqnarray*}
\cos\delta = \Bigl(\frac{\lambda u^{(0)}_{xx}}{2}-\frac{\beta}{4}\Bigr)
\Bigr[(\chi-I_z)^2+\Bigl(\frac{\lambda u^{(0)}_{xx}}{2}-\frac{\beta}{4}
\Bigr)^2\Bigr]^{-1/2}, 
\end{eqnarray*}
and $|i\rangle$ is the eigenvector of the operator $S^z$, $u^{(0)}_{ij}$ are 
spontaneous deformations determined from the condition of the free energy 
density minimum: $F=F_e - T\ln Z$, where $F_e$ is the density of an elastic 
energy, $Z=\sum_M\exp(-E_M/T)$ is the partition function. In our case the 
spontaneous deformations have the form:
\begin{eqnarray*}
u^{(0)}_{xx}=-\frac{\lambda}{E}\cdot\frac{1-2\sigma}{2}, \;\;\; 
u^{(0)}_{zz}=-\frac{\lambda}{E}\cdot\frac{2-\sigma}{2}, \;\;\; u^{(0)}_{xz}=0.
\end {eqnarray*}

Note, that the calculations are carried out in a low temperature limit (the 
temperature is much less than the temperature of the phase transition). In 
this case it is possible to account only the lowest energy level, which as it 
follows from (3), is $E_1$. 

In the basis of eigenfunctions of the operator $H_0(n)$ we shall build the 
Hubbard's operators $X^{MM'}_n\equiv|\Psi_n(M')\rangle\langle\Psi_n(M)|$, 
which describe transitions of a magnetic ion from a state $M'$ to a state 
$M$. These operators are connected with the spin operators by the relations:
\begin{equation}
\begin{array}{c}
S^+_n=\sqrt{2}\cos\delta\cdot(X_n^{10}+X_n^{0-1})+\sqrt{2}\sin\delta\cdot
(X_n^{01}-X_n^{-10}), \\
S_n^- = (S^+_n)^+, \\
S_n^z = \cos 2\delta\cdot(H^1_n-H^{-1}_n) - \sin 2\delta\cdot(H^{1-1}_n - 
H^{-11}_n),
\end{array}
\end{equation}
where $S^{\pm}_n=S_n^x\pm\imath S^y_n$, $H^M_n\equiv X_n^{MM}$ are diagonal 
Habbard's operators. 

In terms of Habbard's operators the one-ion Hamiltonian can be represented in 
the form
\begin{eqnarray*}
H_0(n)=(\vec{E}\vec{H}_n)+\lambda u_{xz}[d_1(X^{01}_n+X^{10}_n)+d_2(X^{0-1}_n 
+ X^{-10}_n)],
\end{eqnarray*}
where we denote:
\begin{eqnarray*}
d_1=\frac{1}{\sqrt{2}}(cos\delta - \sin\delta), \;\;\;\;\; 
d_2=-\frac{1}{\sqrt{2}}(cos\delta + \sin\delta).
\end{eqnarray*}

Further, we represent components of tensor of deformations in the form 
$u_{ij}=u^{(0)}_{ij}+u^{(1)}_{ij}$, where $u^{(1)}_{ij}$ is the dynamic part 
of tensor of deformations describing oscillations of a lattice. Applying to 
$u^{(1)}_{ij}$ the method of harmonic quantization $\cite{land76}$, from the 
Hamiltonian (2) we shall obtain the Hamiltonian, describing processes of 
transformations of magnons to phonons and vice versa:
\begin{equation}
\begin{array}{c}
H_{tr}=\sum_n[\sum_MP_MH^M_n+\sum_{\alpha}P_{\alpha}X^{\alpha}_n], \\
P_{M(\alpha)} = \frac{1}{\sqrt{N}}\sum_{q,\lambda}(b_{q,\lambda}+
b^{\dag}_{-q,\lambda})T^{M(\alpha)}_n(q,\lambda). 
\end{array}
\end{equation}
Here $b_{q,\lambda}(b^{\dag}_{-q,\lambda})$ are rising and lowering operators 
of phonons with polarization $\lambda$, $T^{M(\alpha)}_n(q,\lambda)$ are the 
amplitudes of transformations. 

\section{Spectra of elementary excitations}
As it is known, the spectra of elementary excitations of a system are 
determined by the poles of a Green's function, which we shall define as 
follows:
\begin{eqnarray*}
G^{\alpha\alpha'}(n,\tau;n',\tau')= - \langle\hat{T}X^{\alpha}_n(\tau)
X^{\alpha'}_{n'}(\tau')\rangle ,
\end{eqnarray*}
where $X^{\alpha}_n(\tau)=e^{H\tau}X^{\alpha}_ne^{-H\tau}$, $H=H_{int}+H_{tr}$, 
$\hat{T}$ is a time-ordering operator, and the averaging will be carried out 
with the Hamiltonian $H$. 

As we work in a mean field approximation, for futher calculations we need only 
"transversal" part of the Hamiltonian $H_{int}$, which in terms of the 
Habbard's operators has the form:
\begin{eqnarray*}
H^{\perp}_{int}= - \frac{1}{2}\sum_{n,n';\alpha,\beta}\frac{I(n-n')}{2}\cdot
A_i^{-\alpha}X^{\alpha}_nB^{\beta}_iX^{\beta}_{n'} .
\end{eqnarray*}
Vector-columns $A_i^{\alpha}$ and $B^{\beta}_i$ have the form:
\begin{eqnarray*}
A_1^{\alpha}=
\left (
\begin{array}{c}
2\gamma_{\parallel}(-\alpha) \\
\Gamma_{\parallel}(M)
\end{array}
\right ), \;\;\;\;
A_2^{\alpha}=
\left (
\begin{array}{c}
\gamma_{\perp}^*(\alpha) \\
0
\end{array}
\right ), \;\;\;\;
A_3^{\alpha}=
\left (
\begin{array}{c}
\gamma_{\perp}(-\alpha) \\
0
\end{array}
\right ), \\
B_1^{\alpha}=
\left (
\begin{array}{c}
\gamma_{\parallel}(\alpha) \\
\Gamma_{\parallel}(M)
\end{array}
\right ), \;\;\;\;
B_2^{\alpha}=
\left (
\begin{array}{c}
\gamma_{\perp}(\alpha) \\
0
\end{array}
\right ), \;\;\;\;
B_3^{\alpha}=
\left (
\begin{array}{c}
\gamma_{\perp}^*(-\alpha) \\
0
\end{array}
\right ).
\end{eqnarray*}
The denominator of the Green's function, which satisfies to the Larkin's 
equation $\cite{mit89}$, gives the dispersion equation:
\begin{equation}
\det\Bigr|\delta_{ij}+\frac{I(k)}{2}B^{\alpha}_i\Sigma^{\alpha\alpha'}
A_j^{\alpha'}+\frac{I(k)}{2}\cdot\frac{D_{\lambda}(k,\omega_n)
B^{\alpha}_i\Sigma^{\alpha\alpha'}T^{-\alpha'}(k,\lambda)T^{\gamma}
(-k,\lambda)\Sigma^{\gamma\gamma'}A^{\gamma'}_j}
{1-D_{\lambda}(k,\omega_n)T^{\beta}(k,\lambda)\Sigma^{\beta\beta'}
T^{-\beta'}(-k,\lambda)} \Bigl|=0 .
\end{equation}
In (7) we denote $D_{\lambda}(k,\omega_n)=2\omega_{\lambda}(k)/
[\omega^2-\omega^2_{\lambda}(k)]$ is the Green's function of noninteracting 
phonons, $\omega_{\lambda}(k)=c_{\lambda}k$ is the low of dispersion of 
$\lambda$ - polarized phonons and $c_{\lambda}$ is their velosity, 
$\Sigma^{\alpha\alpha'}(k,\omega)$ is the nonreducible (by Larkin) part, 
$T^{\alpha}(k,\omega)$ are corresponding amplitudes of transformations. In the 
mean field approximation
\begin{eqnarray*}
\Sigma^{\alpha\alpha'}(k,\omega)=\delta_{\alpha\alpha'}b(\alpha)G^{\alpha}_0
(\omega),
\end{eqnarray*}
where $G^{\alpha}_0(\omega)=[\omega + (\vec{\alpha}\vec{E})]^{-1}$ is the 
zero Green's function, $b(\alpha)=\langle\vec{\alpha}\vec{H}\rangle_0$. 

The solutions of the equation (7) determine spectra of quasiphonons and 
quasimagnons, however, as we study the possiblity of stabilization of LMO in 
a $2D$ easy-plane ferromagnetic, i.e. the question of convergence of the 
integral determining fluctuations of the magnetic moment, first of all we 
have to obtain spectra of quasimagnons. 

It is necessary to note, that the dispersion equation (7) is valid for 
arbitrary temperatures and values of constant of OA. 

The solution of the dispersion equation in a low temperature limit allows us 
to determine spectra of quasimagnons:
\begin{equation}
\varepsilon_{\beta}=2\chi, \;\;\; \varepsilon^2_{\alpha}=[E_{10}-I(k)\cdot
(1-\sin 2\delta)]\cdot[E_{10}-I(k)\cdot(1+\sin 2\delta)],
\end{equation}
where $E_{10}=E_1-E_0, \;\; \varepsilon_{\beta}$ is a high-frequency magnon 
branch, $\varepsilon_{\alpha}$ is a low-frequency magnon branch. 

We study the solution (8) in the two limiting cases: the case of small and 
large OA value. 

a). In the case of small OA value, i.e. at $\beta\ll I_0$, a low-frequency 
magnon branch has the spectrum:
\begin{equation}
\varepsilon^2_{\alpha}(k)=(b_0+\alpha k^2)\cdot(b_0+\frac{\beta}{2}+\alpha 
k^2),
\end{equation}
where $b_0=3\lambda^2/4E$ is the parameter of ME exchange, $\alpha =I_0R^2_0$, 
$R_0$ is the radius of interaction. In the case of the absence ME exchange 
we have a spectrum of an easy-plane ferromagnetic:
\begin{equation}
\varepsilon_{\alpha}(k)=\alpha k\sqrt{\frac{\beta}{2}+\alpha k^2} .
\end{equation}

b). If OA value is large, i.e. $(\beta/4)\gg I_0$, the spectrum of 
quasimagnons has the form:
\begin{equation}
\varepsilon^2_{\alpha}(k)=(\frac{\beta}{2}+a_0)\cdot(\frac{\beta}{2}+a_0-
2I_0+2\alpha k^2).
\end{equation}
Here $a_0=\lambda^2(1+\sigma)/2E$ is the parameter of ME exchange. The spectra 
of quasiphonons are also determined by the equation (7), however, as they do 
not influence the size of the fluctuations of the magnetic moment, they are 
irrelevant. For example a spectrum of longitudinal quasiphonons is as 
follows:
\begin{eqnarray*}
\omega^2(k)=\omega^2_t(k)\cdot(1-\frac{a_0}{I_0}). 
\end{eqnarray*}
The similar result is obtained for $t$-polarized quasiphonons. It is obvious, 
that ME interaction merely renormalizes the velocity of the sound. 

\section{The critical temperature}

Consider fluctuations of the magnetic moment, for example, 
$\langle(S^z)^2\rangle$. The most simple way to calculate it is to represent 
the operator $S^z$ in terms of bose operators through bosonisation of the 
Habbard's operators $\cite{val90}$.

Following $\cite{val90}$, we assign to the Habbard's operators $X^{\alpha}_n$ 
pseudohubbard's operators $\tilde{X}^{\alpha}_n$, which are related with 
rising and lowering bose-operators of quasiparticles by the following 
relations:
\begin{eqnarray*}
%\begin{array}{c}
\tilde{X}^{10}_n=(1-a^{\dag}_na_n-b^{\dag}_nb_n)a_n, \;
\tilde{X}^{01}_n=a^{\dag}_n, \;
\tilde{X}^{1-1}_n=(1-a^{\dag}_na_n-b^{\dag}_nb_n)b_n, \;
\tilde{X}^{-11}_n=b^{\dag}_n,\\
\tilde{X}^{0-1}_n=a^{\dag}_nb_n, \;\;\; \tilde{X}^{-10}_n=b^{\dag}_na_n, \;\;\; 
\tilde{H}^0_n=a^{\dag}_na_n, \;\;\tilde{H}^{-1}_n=b^{\dag}_nb_n, \;\;
\tilde{H}^1_n=1-a^{\dag}_na_n-b^{\dag}_nb_n .
\end{eqnarray*}
Here $a$ is the bose operator corresponding to the transitions from a state 
1 to a state 0 and vice versa, $b$ corresponds to the transitions from a state 
1 to a state -1 and vice versa. 

Rewriting the Hamiltonian $H$ in terms of bose operators, and using the fact 
that the one-ion Hamiltonian in the basis of its eigenfunctions becomes 
diagonalize:
\begin{eqnarray*}
H_0(n)=(\vec{E}\vec{H}_n),
\end{eqnarray*}
we write down only two-partical Hamiltonian, that has the form:
\begin{eqnarray*}
H^{(2)}=\sum_k(E_0-E_1-I(k))\cdot a^{\dag}_ka_k+\sum_k(E_{-1}-E_1)\cdot 
b^{\dag}_kb_k-\frac{1}{2}\sin 2\delta\cdot\sum_kI(k)(a_ka_{-k}+a^{\dag}_k
a^{\dag}_{-k}).
\end{eqnarray*}
Diagonalizing the obtained Hamiltonian by the standard $u-v$ transformation:
\begin{eqnarray*}
a^{\dag}_k=u_k\alpha_k+v_k\alpha^{\dag}_{-k}, \;\;\; 
a_k=u_k\alpha_k^{\dag}+v_k\alpha_{-k}, 
\end{eqnarray*}
we receive:
\begin{eqnarray*}
H^{(2)}=\sum_k\varepsilon_{\alpha}\alpha^{\dag}_k\alpha_k+
\sum_k\varepsilon_{\beta}\beta^{\dag}_k\beta_k ,
\end{eqnarray*}
where expressions for bosone energies coincide with (8).

Using the connection of spin operators with the Habbard's operators (5), for 
fluctuations of the magnetic moment we shall receive:
\begin{equation}
\langle(S^z_n)^2\rangle=-\sin 2\delta\langle b^{\dag}_nb_n\rangle+
\frac{1-\sin 2\delta}{2}\cdot\langle a^{\dag}_na_n\rangle +
\frac{1+\sin 2\delta}{2}.
\end{equation}
Summarizing (12) we receive the desired average and
\begin{eqnarray*}
\frac{1}{N}\sum_n\langle b^{\dag}_nb_n\rangle =\frac{1}{(2\pi)^2}\int^{\infty}
_0\frac{kdk}{e^{\varepsilon_{\beta}/T}-1}, \;\;
\frac{1}{N}\sum_n\langle a^{\dag}_na_n\rangle =\frac{1}{(2\pi)^2}\int^{\infty}
_0\frac{kdk}{e^{\varepsilon_{\alpha}/T}-1}, 
\end{eqnarray*}
where $\varepsilon_{\alpha}(k)$ and $\varepsilon_{\beta}(k)$ are determined 
by the expressions (8). The relation (12) is valid for arbitrary values of 
OA constant. As in the previous case we investigate two limiting cases: 
small and larg OA values.

a). In the case of small OA the contribution of a high-frequency branch of 
excitations to fluctuations of the magnetic moment can be neglected. It is 
obvious, that in this case the integral
\begin{eqnarray*}
\int^{\infty}_0\frac{kdk}{e^{\varepsilon_{\alpha}/T}-1}
\end{eqnarray*}
converges on the lower limit ($\varepsilon_{\alpha}(k)$ is determined by the 
formula (9)), hence, in this case there exist a LMO in $2D$ ferromagnetics. 
If we disregard with the ME exchang, the integral diverges on the lower limit, 
that testifies to the absence of LMO. Besides from the condition 
$(1/N)\sum_n\langle(S^z_n)^2\rangle =1$ it is possible to determine the 
temperature of the phase transition. In our case we have:
\begin{eqnarray*}
\frac{1}{(2\pi)^2}\int^{\infty}_0\frac{kdk}{e^{\varepsilon_{\alpha}/
T_c}-1}=1,
\end{eqnarray*}
where $T_c$ is the temperature of the phase transition. Substituting the 
expression (9) for $\varepsilon_{\alpha}(k)$, we obtain the expression for 
$T_c$ :
\begin{equation}
T_c\approx 4\pi\alpha\left[\ln\frac{4\pi\alpha}{\sqrt{b_0\cdot(b_0+\beta/2)}}
\right]^{-1} .
\end{equation}
As it is evident from (13), the temperature of the phase transition is 
determined both by the ME interaction, and OA. But still ME interaction is a 
decisive one, and at $b_0=0, \;\; T_c\rightarrow 0$.

b). In the case of large OA value fluctuations of the magnetic moment is 
equal to 
\begin{eqnarray*}
\frac{1}{N}\sum_n\langle(S^x_n)^2\rangle = 
\frac{1}{(2\pi)^2}\int^{\infty}_0\frac{kdk}{e^{\varepsilon_{\alpha}/
T}-1}, 
\end{eqnarray*}
where $\varepsilon_{\alpha}(k)$ is determined by the formula (11). Bisides 
even in the absence of the ME exchange the integral does not diverge on the 
lower limit (in a spectrum of quasimagnons there is a finite gap at $a_0=0$). 
Evidently, such a behavior of fluctuations is connected with the fact that 
in the case of large OA there may realize the phase with a tensorial order 
parameter $\cite{onuf81}$ (the so-called QU-phase).

As before, we shall determine $T_c$. In our case we have ($\beta/4\gg I_0$):
\begin{equation}
T_c\approx\frac{\sqrt{\Delta}}{\ln(\frac{\Delta}{\pi\xi})}, \;\;\;
\xi\approx\alpha\cdot\beta, \;\;\; \Delta\approx\frac{\beta}{2}\cdot
\Bigr(\frac{\beta}{2}-2I_0+a_0\Bigl).
\end{equation}

\section{Conclusion}
The carried out investigations show, that the account of the ME interaction 
results in the stabilization of the LMO in a $2D$ ferromagnetic. From the 
formula (13) it follows, that in a low-anisotropic FM the value of $T_c$ 
is determined by the ME gap of a quasimagnon spectrum.

In the case of large OA the influence of ME interaction on the establishing 
of LMO is not so essential, since in such systems there can exist a 
magnetically  ordered state with a tensorial order parameter (the so-called 
quadrupole phase) $\cite{ivas95}$. Temperature of transition, as follows 
from (14) is determined mainly by the constant of OA, and the ME interaction 
only renormalizes it. 

One of the authors (S.D.) thanks International Soros Program of support of 
education in the field of exact sciences, Grant Nr {\bf SU072163}.

\end{document}